# Who Tracks Who? A Surveillance Capitalist Examination of Commercial Bluetooth Tracking Networks

Hongrui Jin

kevin.jin@student.uva.nl

**Abstract**

Object and person tracking networks powered by Bluetooth and mobile devices have become increasingly popular for purposes of public safety and individual concerns. This essay examines popular commercial tracking networks and their campaigns from Apple, Samsung and Tile with reference to surveillance capitalism and digital privacy, discovering the hidden assets commodified through said networks, and their potential of turning users into unregulated digital labour while leaving individual privacy at risk.

# Introduction

## Keeping Track: From Things to People

During the coronavirus pandemic, many institutions have developed contact-tracing methods to monitor the spread of the disease amongst their population. One of the common methods is to use Bluetooth broadcasting and connection to acquire proximity data from mobile devices from individuals. This technology was by no means new; companies like Apple have been using their tracking networks to track devices and items adjacent to specific devices such as smartphones, tablets and computers. Other tracking networks include Samsung's SmartThings Find and Tile. These networks operate very similarly: devices and items broadcast their presence to surrounding supported devices via Bluetooth, while the latter recognises the device or item and upload their location to a cloud service (Greenberg 2019). The tracked item does not require long distance transmissions as it only needs to sends signals to nearby devices with Internet access. Since the update is passive, offline tracking is possible and supporting devices are cheaper to make comparing to active trackers with GPS and cellular features. However, tracking networks are always exclusive to a certain brand. The FindMy app is only available on Apple devices, and Samsung SmartThings only supports Galaxy phones. Companies claim their networks to be secure, private and anonymous (Apple Inc. n.d.a), yet stalking and other exploits have been repeatedly reported (Buffamonte 2022; Heinrichs 2022). Although these networks claim to be secure, they utilise users private information as the base-stone of their business.

## Defining Privacy

Companies are always seen advertising their tracking networks as "safe" and "private". These terms, however, are vague and hard to quantify. Alan Westin in his work *Privacy and Freedom*, defined privacy as the power of an individual to determine "when, how, and to what extent information about them is communicated to others" (Westin and Solove 2018). Building from Westin, Hung and Wong (2009, 157-58) further defined digital privacy into 3 categories: information privacy (to determine by themselves what and how their information is shared), communication privacy (free from eavesdropping and interception) and individual privacy (the right to be left alone without interruption). This newer definition not only acknowledges the discreet nature of personal information, but also notes the individual's choice to temporarily withdraw from the society, both of which presents greater relevance to modern media technologies. Therefore, Hung and Wong's definitions serves as qualitative criteria that allows for tracking network to be analysed in terms of its privacy protection. What's more, such privacy is also transformed into a profitable service, resembling a particular form of information capitalism: surveillance capitalism.

**Make Use of Users: Surveillance Capitalism**

Coined by Shoshana Zuboff first in 2014, surveillance capitalism refers to the monetisation of personal information acquired through means of surveillance and data extraction (Zuboff 2014). The information extracted and generated from users are therefore called "surveillance assets", and could be later transformed into capital. In comparison to information capitalism, surveillance capitalism earned its name from the elaborate and timely information it acquires via various channels from smartphones to fitness trackers. Search engines and targeted advertisements, such as Google, is an example of explicit surveillance capitalism: users' search history is stored and analysed by Google as the surveillance asset, and was transformed into capital by advertisers paying Google to match their campaigns to users with certain traits. When examining tracking networks under a surveillance capitalist scope, the surveilled asset is not immediately clear. Companies seemingly received no benefits by promising "end to end encryption" and anonymity. Meanwhile, the entire network alongside its users were capitalised to other manufacturers as well as consumers in two ways: manufacturers are required a licensing fee to participate in such network, customers pay for trackers such as AirTags. In layman's terms, this network allows companies to sell user locations and their surroundings data barely at their consent. Further textual observation reveals companies always associate privacy with anonymity, though they very much differ.

**Recognising Digital Labour**

Since tracking networks and supporting hardwares are commercialised, it is important to question what exactly added value to this service. Fuchs and Sevignani (2013, 237) describes digital labour as the work incorporating human experience with digital media to create products. While Fuchs and Sevignai applied this definition to users of Facebook, it could be argued that this Marxist framework of justifying labour can also be relevant to other media infrastructures, as long as they commodify user communications. In the example of a surveillance capitalist business model, digital labour *is* the infrastructure: unlike Google or Facebook, there will be no service without large amounts of users. Examining users in tracking networks as digital labour invites a critique of how customers were recruited and exploited at the very foundation of surveillance capitalism.

**Guiding Questions**

Interrogating tracking network services with aforementioned concepts, two key inquiries arise:

> What is commodified behind the tracking network services?
>
> How do companies describe security and privacy to their customers?

This essay asserts that although major crowdsourced tracking networks are advertised safe and convenient, its surveillance capitalist nature reaps profit by exploiting user privacy under a lack of regulation, all at a scale of hundred of millions of devices (Apple Inc. n.d.a).

**<u>Methodology</u>**

This study first identified leading tracking network providers with reference to industry reports, revealing Apple, Samsung and Tile as market leaders. These major companies have a wider reach to global customers, making them suitable to observe social and cultural impacts at a larger scale. After selecting market leaders, advertisement materials of these tracking networks were collected from their websites and video platform (YouTube), as these contents are more likely to be viewed by the public. While some companies did not have a dedicated website for their network, information on such services were alternatively found on webpages for tracker tags and support articles. A total of 10 videos [Apple (n=4), Samsung (n=2), Tile (n=4)], 10 official webpages and articles [Apple (n=6), Samsung (n=2), Tile (n=2)], along with the Apple "Find My" application user interface(UI) were investigated. A content analysis was conducted on the videos following a framework by Priest (1996, 66-67), achieved by systematically analysing visual and narrative elements according to their cultural representation, while the webpages and application UI underwent detailed textual and visual examination for a feature comparison while revealing their privacy treatment. Other application interfaces were not examined due to limited device support, yet they offer very similar functionalities. It is worth noting that technical details of tracking networks are usually proprietary, therefore an analysis of the tracking mechanism could, at best, still be reliant on the disclosure of companies.

## Analysis

### What is Sold?

Sourcing from websites of major tracking network providers, the service itself is usually commercialised through smart trackers (smart tags), or as a feature on other smart devices. All three companies have produced introductory pages for their smart trackers. In their overviews, companies started with the benefits of locating nearby items easily with audio and visual guides from the tracker. When demonstrating tracker usage in longer distances, companies supports its features by introducing tracking networks, which usually consist of current owners of their branded devices. On one hand this feature adds attractiveness and value to the product; on the other hand, this added value was only made possible by incorporating multiple millions of devices continuously broadcasting, listening and uploading Bluetooth packages and location data (Greenberg 2019). Servers analyse uploaded data from users, cross-reference them with the signature of a given lost device, then send users its last seen location if matched. Samsung and Tile confirms this mechanism without disclosing the amount of users in their network(Samsung n.d.; Tile n.d.), yet Apple noted "hundred of millions of friends" could help locating (Apple Inc. n.d.a). This was seen printed on the webpage in a significantly larger bold font, followed by a section explaining the functionality in detail. In its promotional video *How It Works*, Tile claimed it had "the world's largest lost-and-found community" that locates "more than 4 million items everyday" (Tile 2018). Such design choices highlighted that trackers were created to make use of a broader network based on devices from other users. Since remote tracking was promoted as a feature and facilitated under similar network designs by every dominant provider, it is clear that the data collected by users' devices was transformed into a service and commodified as network-based trackers along with other devices that enabled networked tracking, such as phones and tablets.

**Unpaid Labour**

According to the definition by Fuchs and Sevignani (2013, 237), data generated by users that could turn into capital via digital platforms are considered "digital work". In a tracking network, providers do not always directly track users and items. Rather, it is the other users' devices that actively listens to Bluetooth broadcasts and eventually locate any lost items. These networks are commonly introduced as a service based on an unseen infrastructure on the "cloud". However, unlike other services or applications, these networks can only function with other users participating. Users can still use Google if no one else around them is using it; however, there will be no tracking networks if there are no users nearby. The quality of a purchased service is therefore based dominantly on its users, rather than the company or their algorithms. This indicates that users are paying companies to facilitate other customers, which were not rewarded. Users enrolled in tracking networks are therefore entitled digital labourers, yet their work is unpaid since no current regulation applies. Moreover, companies often create a "straw-man" situation, awarding users with services they already deserve as a return for enrolling. For example, offline tracking is only available when users opt in Apple's Find My network, whereas Tile trackers can only be used when user agrees to join its tracking community.

**The Invisible Panorama**

While advertising on the usability and reliability of tracking networks, the data gathering process was far from being advertised and usually hidden from the users' sight. Apple's Find My network, for example, automatically enrols all supported devices as long as users agree to locate their own devices (Apple Inc. n.d.d). Samsung and Tile took a similar approach to network their devices and applications (Thomas 2020). When using the Find My app, users are only prompted to review permission on general location services, yet no dedicated permission was required to listen to or upload surrounding device data, resulting in poor user unawareness of such data extraction. As a result, millions of users' device details, combined with any digital surrounding that has Bluetooth broadcasting could be recorded and uploaded, all in a matter of seconds (Heinrich et al. 2021, 3-5). Through operating device with minimal consent and integrated far-reaching data collection, tracking networks effectively become an unnoticed grid for surveillance. This network extends the already ubiquitous surveillance of companies that converts data to hardware sales and subscriptions. But users, best summarised by Shoshana Zuboff, "are to be harvested for behavioural data" (Zuboff et al. 2019). It is important to point out that the data harvested does not belong to any company; yet they are commodified regardless simply because users did not claim them, and regulations were not yet in place (Zuboff 2014). These factors combined spawned surveillance capitalist tracking platforms that allows companies to extract users' private data in the name of providing convenience.

**Sugarcoated Security**

It is particularly crucial to examine the discourse of security and privacy when it comes to advertising tracking networks and accessories, since their key feature is loss and theft prevention. It is no surprise that major providers offers extra features to minimise loss: Apple and Samsung offers offline finding and precision finding for additional tracking accuracy. On its webpage, Apple demonstrated having every device and item shown on a map in the Find My interface (Apple Inc. n.d.b). This visual demonstration pictures one's devices and belongings in a panoramic view, according to Foucault (1980), this vision communicates a sense of power and control by gaining knowledge, in this case, by knowing the current locations of one's valuables. Samsung and Tile offered a similar demonstration, depicting several items shown on a map within their interface applications. This reinforces the notion of control and the power of knowing, persuading customers that trackers could assist them keep track of things. In other promotional videos, companies envisioned tracking tags to be attached to keys (Apple UAE 2021), camera, wallets (Tile 2017), pets (Samsung Gulf 2021), and interestingly in a 2017 commercial, a stuffed toy panda dear to a girl (Tile 2017). These videos signified the trackers potential of securing one's priced possession, inviting customers to, similarly, "tag" important items with their trackers, whether such importance was monetary or sentimental. Tile's slogan, "together we find" attempts to convince customers that their items would be secure under the lookout from its networked community. However, the cost for this advertised "security", a continuous scan and broadcasting of Bluetooth signals, was approached and disclosed in more ambiguous terms, or sometimes intentionally evaded. Although the company usually explains privacy protocols in details (see Apple Inc. n.d.c), Apple simply announced their tracking network to be "secure", "anonymous" and "encrypted" without providing any detailed technical explanations (Apple Inc. n.d.b). Tile described the process solely as "anonymous" (Tile n.d.), while Samsung did not address data protection at all, but addressed in its footnotes the conditions of using offline tracking, requiring enrolment and sign-in (Samsung n.d.). The footnotes were printed in an extremely small and light font, making it hardy legible unless closely observed. The extensive use of arbitrary and oftentimes inadequate descriptors from leading companies revealed a potential lack of data protection in tracking networks. Moreover, cases of stalking using trackers have been reported due to the exclusiveness of each companies' tracking network (Buffamonte 2022). Both Apple and Samsung specified that their tracking network only supports devices of their own brand, so even though Apple provides anti-stalking features against unidentified Apple trackers (Apple Inc. n.d.a), it could not detect a Samsung tracker since there is no universal standards for trackers to communicate with networked devices.

**Conflicting Privacy**

A closer look at the tracking network reveals its users being remotely surveilled by repeatedly submitting Bluetooth and GPS data, only to receive locations and directions to their items (Haskell-Dowland 2021). This is typical of an asymmetric surveillance system, otherwise known as a *panopticon,* where individuals are watched without knowing what or how exactly they were surveilled (McMullan 2015). The surveillance capitalist nature of these platforms ensures all extracted data are ultimately centralised and analysed under the name of proprietary information, further privatising individual secrecy. Revisiting the definition of privacy coined by Hung and Wong (2009), an centralised asymmetric surveillance system fundamentally undermines privacy in various ways: information privacy is merely respected superficially, for users have no further control the sharing of their information once uploaded; even when guaranteed anonymity and encryption, individual privacy is exploited, for users can no longer be truly left alone since other users' devices can also detect their presence, not to mention intercepting and exposing their digital signature (Hay and Harle 2009). An analogy would be taking photos: When a photo of you is already taken, the photographer no longer matters. In a tracking network, millions of devices take invisible "photos" of their digital surroundings, then uploaded to manufacturers for commodification and analysis. Individual private information, when accumulated and assembled, becomes the asset for surveillance capitalism (Zuboff et al. 2019).

**Conclusion**

With reference to surveillance capitalism and digital privacy, advertisements and articles were analysed to discover the hidden assets behind tracking networks, and companies notion of privacy when it comes to a safety oriented product that was based on privacy exploits. The development of crowdsourced tracking networks has certainly became beneficial for users looking for their devices or belongings. While the advertised notion of looking after each other in a networked and secure community is seemingly attractive, a detailed analysis of leading tracking service (Apple, Samsung and Tile) and their advertising discourse unearthed some fundamental flaws of the current centralised and asymmetrical approach to creating a tracking network. Enrolment turns users and their devices into companies' unpaid and unregulated digital labour, leading to data extraction with minimal user consent, which in turn feeds companies more data and sustains the tracking network itself. Users data was transformed into capital by selling tracker tags and services that inherently benefitted from this crowdsourced infrastructure. Additionally, despite the claims of security and privacy by service providers, a continuous broadcasting and listening system makes users always visible, either through their own devices or through other individuals' enrolled hardwares, while a lack of regulation allows companies to accumulate Bluetooth data from users without their discretion, leaving space for contact-based tracking and big data composition.

It should be noted that even though this essay pointed out several flaws of commercialised tracking networks, the author still believes that tracking networks is an important sector amongst the internet of things, and affirms that by implementing a balanced, decentralised structure instead of an asymmetrical, centrally computed platform, as well as using an on-demand tracking mechanism instead of continuous listening and broadcasting, combined with comprehensive data protection regulations for Bluetooth communications, such networks could be significantly improved.